\documentclass[smallextended]{svjour3}
\usepackage{graphicx, amsmath, amssymb}
\usepackage[caption=false]{subfig}

\newcommand{\ket}[1]{{\left| #1 \right\rangle}}
\newcommand{\ketbra}[2]{{\left| #1 \middle\rangle \middle \langle #2 \right|}}

\journalname{Quantum Inf Process}

\begin{document}

\title{Equivalence of Szegedy's and Coined Quantum Walks}

\author{Thomas G.~Wong}

\authorrunning{T.~G.~Wong}

\institute{T.~G.~Wong \at
	   Department of Computer Science, University of Texas at Austin, 2317 Speedway, Austin, Texas 78712, USA \\
	   Currently at Department of Physics, Creighton University, 2500 California Plaza, Omaha, NE 68102, USA \\
	   \email{thomaswong@creighton.edu}
}

\date{Received: date / Accepted: date}

\maketitle

\begin{abstract}
	Szegedy's quantum walk is a quantization of a classical random walk or Markov chain, where the walk occurs on the edges of the bipartite double cover of the original graph. To search, one can simply quantize a Markov chain with absorbing vertices. Recently, Santos proposed two alternative search algorithms that instead utilize the sign-flip oracle in Grover's algorithm rather than absorbing vertices. In this paper, we show that these two algorithms are exactly equivalent to two algorithms involving coined quantum walks, which are walks on the vertices of the original graph with an internal degree of freedom. The first scheme is equivalent to a coined quantum walk with one walk-step per query of Grover's oracle, and the second is equivalent to a coined quantum walk with two walk-steps per query of Grover's oracle. These equivalences lie outside the previously known equivalence of Szegedy's quantum walk with absorbing vertices and the coined quantum walk with the negative identity operator as the coin for marked vertices, whose precise relationships we also investigate.
	\keywords{Szegedy's quantum walk \and Coined quantum walk \and Spatial search \and Query model}
	\PACS{03.67.Ac, 02.10.Ox}
\end{abstract}


\section{Introduction}

Quantum walks are one of the primary tools for developing quantum algorithms. For example, they have resulted in quantum algorithms for searching \cite{SKW2003,CG2004}, element distinctness \cite{Ambainis2004}, triangle finding \cite{Magniez2007}, and boolean formula evaluation \cite{FGG2008,Ambainis2010}. They also speed up backtracking algorithms \cite{Montanaro2015} and are universal for quantum computation \cite{Childs2009,Lovett2010,Childs2013}. Several reviews of quantum walks are available, including \cite{Kempe2003,Ambainis2003,Venegas2012}.

There are several definitions of quantum walks, depending on the scheme for quantizing a classical random walk. In discrete-time, the most popular models are Szegedy's quantum walk and coined quantum walks. Szegedy's quantum walk \cite{Szegedy2004} arises from quantizing a classical random walk or Markov chain. In his scheme, the walk occurs on the edges of the bipartite double cover of the original graph, and the evolution is defined by two reflection operators.

Coined quantum walks, on the other hand, walk on the vertices of the original graph. They predate Szegedy's quantum walk and even provided inspiration for Szegedy's work \cite{Szegedy2004}. Coined quantum walks began with Meyer \cite{Meyer1996a,Meyer1996b}, who investigated them in the context of quantum cellular automata. He showed that an internal spin degree of freedom could be included to make the evolution nontrivial, and the resulting dynamics are a discretization of the Dirac equation. Aharonov \textit{et al.}~\cite{Aharonov2001} framed this evolution as a quantum walk, renaming the internal state a coin, in reference to a coin flip in classical random walks. The quantum walk is then defined by a coin flip followed by a shift or hop to adjacent vertices.

Under certain conditions, it is known from Fact 3.2 of \cite{Magniez2012} that one step of Szegedy's quantum walk is equivalent to two steps of the coined quantum walk. This fact is stated without proof, however, and the precise relationships between the individual operators are not given. Some details are described in lecture notes \cite{Whitfield2012}, but again the exact relationships between individual operators are not explored. In this paper, we amend this oversight, explicitly showing that Szegedy's first reflection operator is equal to the ``Grover diffusion'' coin flip of the coined quantum walk, while Szegedy's second reflection operator is equal to the combination of a ``flip-flop'' shift, Grover diffusion coin flip, and another flip-flop shift. These relationships between the operators also hold for the seminal search schemes, where Szegedy's quantum walk quantizes a Markov chain with absorbing marked vertices \cite{Szegedy2004}, and the coined quantum walk uses the Grover diffusion coin for unmarked vertices and the negative identity for marked vertices \cite{SKW2003}.

A modified version of Szegedy's quantum walk that includes complex phases was recently shown to be equivalent to a modified coined quantum walk on a multigraph, under certain conditions \cite{Portugal2016b,Portugal2017}. In this scenario, Szegedy's first reflection operator is equal to the coin flip, and Szegedy's second reflection operator is equal to the shift. Hence, one step of Szegedy's quantum walk is equal to \emph{one} step of the coined. In this paper, however, we focus on the standard definitions of the quantum walks, where one step of Szegedy's quantum walk is equal to \emph{two} steps of the coined.

Recently, Santos \cite{Santos2016} proposed two alternative methods for searching with Szegedy's quantum walk that, rather than using absorbing vertices, utilizes oracles similar to the one from Grover's algorithm \cite{Grover1996}. She proved that her first algorithm is equivalent to Szegedy's algorithm with absorbing vertices for a certain class of graphs. That is, for these graphs, Szegedy's search algorithm with absorbing vertices can be expressed in terms of a Grover-type oracle. For her second algorithm, Santos showed that it achieves a better success probability when searching the complete graph, indicating that search with Grover-type queries can be superior. Furthermore, the improved success probability of this algorithm can be used to achieve perfect state transfer \cite{Stefanak2016}.

This raises the question of whether Santos's schemes, based on Szegedy's quantum walk, are equivalent to coined quantum walks. Santos's algorithms are not encompassed within the framework of Fact 3.2 of \cite{Magniez2012}, so in this paper, we give the coined quantum walk search algorithms that are equivalent to Santos's two search algorithms. We prove that her first algorithm is equivalent to a coined quantum walk with one walk-step per query to a Grover-type oracle \cite{AKR2005}, and her second algorithm is equivalent to a coined quantum walk with two walk-steps per query to a Grover-type oracle \cite{Wong11}. Both of these are established concepts in coined quantum walks. Then we are able to explain Santos's results using the large body of research on coined quantum walks and expand upon her conclusions.

In the next section, we define Szegedy's quantum walk, review how it searches with absorbing vertices, and then define Santos's search schemes. Then in Section III, we follow the same outline as Section II, except with equivalent coined quantum walks. So we define the coined quantum walk and prove its equivalence to Szegedy's by giving the explicit relationships between the operators. Then we explain its seminal search algorithm and prove its equivalence to Szegedy's search with absorbing vertices. Finally, we give the two coined quantum walks that are equivalent to Santos's Szegedy-based search algorithms with Grover-type oracles. This allows us to understand Santos's results using the vast literature on coined quantum walks.


\section{Szegedy's Quantum Walk}

\subsection{Definition}

\begin{figure}
\begin{center}
	\subfloat[]{
		\includegraphics{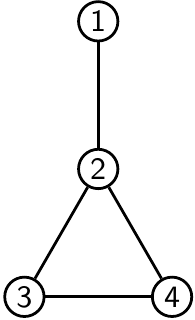}
		\label{fig:graph}
	} \quad
	\subfloat[]{
		\includegraphics{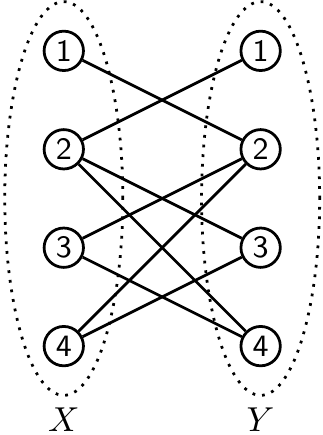}
		\label{fig:graph_szegedy}
	}
	\caption{(a) An irregular graph of $N = 4$ vertices, and (b) its bipartite double cover.}
\end{center}
\end{figure}

We assume throughout the paper that the classical random walk we are quantizing is unbiased, meaning it occurs on an undirected and unweighted graph. For example, for the graph in Fig.~\ref{fig:graph}, a classical random walker at vertex 2 has probability $1/3$ each of jumping to vertices 1, 3, and 4. Similarly, a walker at vertex 3 has probability $1/2$ each of jumping to vertices 2 and 4.

Szegedy's quantum walk occurs on the edges of the bipartite double cover of the original graph. Mathematically, if the original graph is $G$, then its bipartite double cover is the graph tensor product $G \times K_2$, where $K_2$ is the complete graph of two vertices. This duplicates the vertices into two partite sets $X$ and $Y$. A vertex in $X$ is connected to a vertex in $Y$ if and only if they are connected in the original graph. For example, the bipartite double cover of Fig.~\ref{fig:graph} is shown in Fig.~\ref{fig:graph_szegedy}. Note if the number of edges in the original graph is $|E|$, the number of edges in its bipartite double cover is $2|E|$. Since the quantum walk occurs on the edges of the bipartite double cover, the Hilbert space of the walk is $\mathbb{C}^{2|E|}$. We denote a walker on the edge connecting $x \in X$ with $y \in Y$ as $\ket{x,y}$. Then the computational basis is
\[ \{ \ket{x,y} : x \in X, y \in Y, x \sim y \}, \]
where $x \sim y$ denotes that vertices $x$ and $y$ are adjacent.

Szegedy's walk is defined by repeated applications of
\[ W = R_2 R_1, \]
where
\begin{gather*}
	R_1 = 2 \sum_{x \in X} \ketbra{\phi_x}{\phi_x} - I, \\
	R_2 = 2 \sum_{y \in Y} \ketbra{\psi_y}{\psi_y} - I
\end{gather*}
are reflection operators defined by
\begin{gather*}
	\ket{\phi_x} = \frac{1}{\sqrt{\deg(x)}} \sum_{y \sim x} \ket{x,y}, \\
	\ket{\psi_y} = \frac{1}{\sqrt{\deg(y)}} \sum_{x \sim y} \ket{x,y},
\end{gather*}
where $\text{deg}(x)$ is the degree of vertex $x$ (\textit{i.e.}, the number of neighbors of $x$), and $y \sim x$ sums over the neighbors of $x$. Note that $\ket{\phi_x}$ is the equal superposition of edges incident to $x \in X$, and $\ket{\psi_y}$ is the equal superposition of edges incident to $y \in Y$. As proved in \cite{Wong25} and in congruence with the ``inversion about the mean'' of Grover's algorithm \cite{Grover1996}, the reflection $R_1$ goes through each vertex in $X$ and reflects the amplitude of its incident edges about their average amplitude, and $R_2$ similarly does this for the vertices in $Y$.

For example, for the bipartite double cover in Fig.~\ref{fig:graph_szegedy}, say the amplitude of edge $\ket{2,1}$ is $c_{2,1}$, the amplitude of edge $\ket{2,3}$ is $c_{2,3}$, and the amplitude of edge $\ket{2,4}$ is $c_{2,4}$. They are all incident to vertex $2 \in X$, and their average is
\[ \bar{c}_2 = \frac{c_{2,1} + c_{2,3} + c_{2,4}}{3}. \]
When $R_1$ is applied, each of the three amplitudes are inverted about this mean, so
\begin{gather*}
	c_{2,1} \rightarrow 2 \bar{c}_2 - c_{2,1}, \\
	c_{2,3} \rightarrow 2 \bar{c}_2 - c_{2,3}, \\
	c_{2,4} \rightarrow 2 \bar{c}_2 - c_{2,4}.
\end{gather*}
$R_1$ does a similar inversion at each vertex in $X$, and $R_2$ does this for vertices in $Y$.


\subsection{Search with Absorbing Vertices}

\begin{figure}
\begin{center}
	\subfloat[]{
		\includegraphics{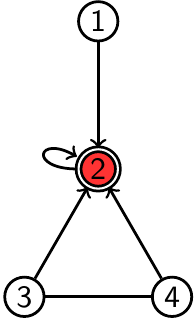}
		\label{fig:graph_marked}
	} \quad
	\subfloat[]{
		\includegraphics{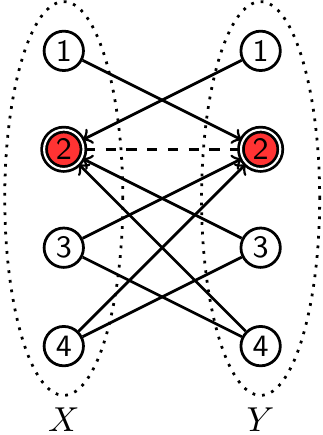}
		\label{fig:graph_szegedy_marked}
	} \quad
	\subfloat[]{
		\includegraphics{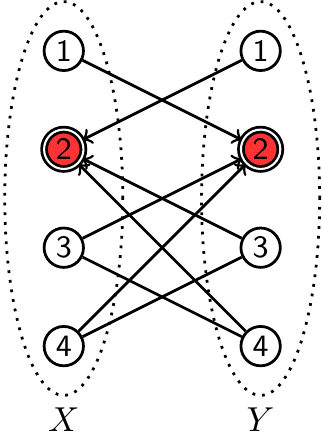}
		\label{fig:graph_szegedy_marked_dropped}
	}
	\caption{(a) Search for vertex $2$ on an irregular graph of $N = 4$ vertices, (b) its bipartite double cover, and (c) the nonzero edges in Szegedy's quantum walk search algorithm.}
\end{center}
\end{figure}

To search for a marked vertex in a graph with a classical random walk, one randomly walks until a marked vertex is found, and then the walker stays put at the marked vertex. Given this procedure, marked vertices act as absorbing vertices, as illustrated in Fig.~\ref{fig:graph_marked}.

Szegedy's quantum walk searches by quantizing this random walk with absorbing vertices, and the resulting bipartite double cover is shown in Fig.~\ref{fig:graph_szegedy_marked}. As shown in \cite{Wong25}, the dashed line connecting vertex $2$ in $X$ and $Y$ can be ignored since it has zero amplitude throughout the evolution, so the effective graph is Fig.~\ref{fig:graph_szegedy_marked_dropped}. Then the search is performed by repeatedly applying
\[ W' = R_2' R_1', \]
where the prime distinguishes that we are searching for absorbing vertices. At unmarked vertices, $R_1'$ and $R_2'$ act identically to $R_1$ and $R_2$, respectively, by inverting the amplitudes of the edges around their average at each vertex. At marked vertices, however, $R_1'$ and $R_2'$ act differently \cite{Wong25}, flipping the signs of the amplitudes of all incident edges.

This seminal search scheme has been investigated for a variety of graphs, including the one-dimensional periodic lattice or cycle \cite{Wong25,Santos2010b} and the complete graph \cite{Santos2010a}. Szegedy also investigated his search scheme for general symmetric and vertex-transitive graphs with a unique marked vertex \cite{Szegedy2004}.


\subsection{Search with Grover's Oracle}

In Grover's algorithm \cite{Grover1996}, the system evolves by repeatedly applying two reflections: The first reflects the state through the marked vertex, and the second reflects across the initial uniform state. This first reflection acts as an oracle query $Q$, and it negates the amplitude at the marked vertex. That is,
\[ Q \ket{x} = \left\{ \begin{array}{rl}
	-\ket{x}, & x\ {\rm marked} \\
	\ket{x}, & x\ {\rm unmarked} \\
\end{array} \right.. \]

Motivated by this, Santos \cite{Santos2016} defined Grover-type oracles in Szegedy's scheme. In the bipartite double cover (\textit{c.f.}, Fig.~\ref{fig:graph_szegedy_marked_dropped}), there are marked vertices in each partite set $X$ and $Y$. So we get two Grover-type oracles, one for each set:
\begin{gather*}
	Q_1 = \left( I_N - 2 \!\!\!\!\!\! \sum_{x \in {\rm marked}} \!\!\!\!\!\! \ketbra{x}{x} \right) \otimes I_N, \\
	Q_2 = I_N \otimes \left( I_N - 2 \!\!\!\!\!\! \sum_{y \in {\rm marked}} \!\!\!\!\!\! \ketbra{y}{y} \right).
\end{gather*}
$Q_1$ flips the sign of an edge if its incident to a marked vertex in $X$, and $Q_2$ acts similarly, except it flips the sign of an edge if its incident to a marked vertex in $Y$.

Using these Grover-type queries, Santos introduced several Szegedy-based schemes for searching. The first algorithm incorporates queries in both the $X$ and $Y$ partite sets, and it repeatedly applies
\[ W_{q1} = R_2 Q_2 R_1 Q_1. \]
The second algorithm only applies queries to vertices in $X$, repeatedly applying
\[ W_{q2} = R_2 R_1 Q_1. \]
Santos also proposed using $R_2 Q_1 R_1 Q_1$, but since $Q_1 R_1 Q_1 = R_1$, this is equivalent to $W = R_2 R_1$, so it is a pure walk that does not search. Similarly, Santos numerically showed that $Q_1 R_2 Q_1 R_1$ causes an initially uniform state to evolve negligibly. Thus, we only focus on $W_{q1}$ and $W_{q2}$ here.

Santos \cite{Santos2016} showed that the first operator $W_{q1}$ is sometimes equal to Szegedy's original search operator $W'$, meaning Szegedy's search can be interpreted as a Grover-type query algorithm in those cases. In particular, she showed this to be true for strongly regular graphs with one marked vertex \cite{Santos2016}, which are simple enough that the evolution occurs in a 4D subspace. Then it can be shown that $t$ applications of $W_{q1}$ on the initial state of the search algorithm is exactly equal to $t$ applications of $W'$.

For her second algorithm, Santos showed that $W_{q2}$ outperforms Szegedy's $W'$ when searching the complete graph. In particular, if the complete graph has $N$ vertices, and $k$ of them are marked, then each operator reaches the following success probability $p_*$ after $t_*$ applications of the respective operators:
\[ \begin{array}{lll}
	W': & p_* = \frac{1}{2}, & t_* = \frac{\pi}{4} \sqrt{\frac{N}{2k}} \\
	W_{q2}: & p_* = 1, & t_* = \frac{\pi}{4} \sqrt{\frac{N}{k}}.
\end{array} \]
That is, with Szegedy's original absorbing operator $W'$, the success probability reaches $1/2$, so we expect to repeat the algorithm twice before finding the marked vertex, on average, leading to an overall runtime that is a factor of $\sqrt{2}$ slower than Santos's alternative operator $W_{q2}$. In addition, Szegedy's absorbing operator $W'$ can be interpreted as making two queries per iteration, one in $X$ and another in $Y$, whereas Santos's operator $W_{q2}$ only makes one query in $X$ per iteration. This observation further makes Szegedy's search algorithm slower than Santos's in the number of queries.

Next, we give the coined quantum walks that are equivalent to Szegedy's quantum walk, Szegedy's search algorithm with absorbing vertices, and Santos's algorithms with Grover-type oracles. We will see that Santos's schemes are equivalent to established coined quantum walks, and so those results carry into Santos's framework and provide extensions to her work.


\section{Equivalent Coined Quantum Walks}

\subsection{Definition}

\begin{figure}
\begin{center}
	\subfloat[]{
		\includegraphics{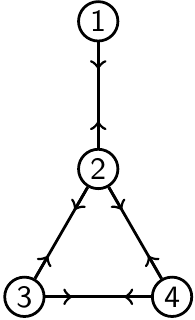}
		\label{fig:graph_coined}
	} \quad
	\subfloat[]{
		\includegraphics{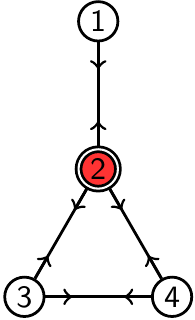}
		\label{fig:graph_coined_marked}
	}
	\caption{(a) A coined quantum walk on an irregular graph of $N = 4$ vertices, and (b) search for vertex $2$.}
\end{center}
\end{figure}

A coined quantum walk jumps on the vertices of the original graph, not the bipartite double cover. It has an additional coin degree of freedom, however, indicating which direction the walker points. This is illustrated in Fig.~\ref{fig:graph_coined}, where each vertex has directions associated with it. For example, a particle at vertex $2$ can point towards vertices $1$, $3$, and $4$ in superposition. Note the number of directions in which a walker at vertex $v$ can point is $\text{deg}(v)$, so the Hilbert space is $\mathbb{C}^{\sum_v \text{deg}(v)}$. If the particle is at vertex $a$ and points towards vertex $b$, we write the state as $\ket{a,b}$. Then the computational basis is
\[ \left\{ \ket{a,b} : a,b \in \{1, \dots, N\}, a \sim b \right\}, \]
The quantum walk is defined by repeated applications of
\[ U = S C, \]
where $C$ is the coin flip that shuffles the internal state of the particle at each vertex, and $S$ is the shift that causes the particle to hop. In this paper, we consider the most common choices for both. For the coin, we use Grover's diffusion coin, which is the permutation-symmetric operator that is furthest from the identity operator \cite{Moore2002}, and it is defined to be
\[ C = 2 \sum_{a = 1}^N \ketbra{s_a}{s_a} - I, \]
where
\[ \ket{s_a} = \frac{1}{\sqrt{\deg(a)}} \sum_{b \sim a} \ket{a,b} \]
is the state of a particle at vertex $a$ uniformly pointing towards each of its neighbors. Then for each vertex $a$, $C$ reflects the internal coin state across the equal superposition $\ket{s_a}$. As shown in \cite{Wong23} and in congruence to the ``inversion about the mean'' of Grover's algorithm \cite{Grover1996}, for each vertex $a$, $C$ inverts the amplitude of each coin state at $a$ about the average amplitude of coin states at $a$. 

For example, for the coined quantum walk in Fig.~\ref{fig:graph_coined}, consider a particle at vertex $2$. Say it points to vertices $1$, $3$, and $4$ with respective amplitudes $c_{2,1}$, $c_{2,3}$, and $c_{2,4}$. The average of these amplitudes is
\[ \bar{c}_2 = \frac{c_{2,1} + c_{2,3} + c_{2,4}}{3}. \]
When $C$ is applied, each of the three amplitudes are inverted about this mean, so
\begin{gather*}
	c_{2,1} \rightarrow 2 \bar{c}_2 - c_{2,1}, \\
	c_{2,3} \rightarrow 2 \bar{c}_2 - c_{2,3}, \\
	c_{2,4} \rightarrow 2 \bar{c}_2 - c_{2,4}.
\end{gather*}
The coin $C$ inverts about the mean at each vertex.

For the shift $S$, we use the flip-flop shift, which causes the particle to hop and then turn around. For example, a particle at vertex $a$ pointing to vertex $b$ jumps to vertex $b$ and points at vertex $a$, so $S\ket{a,b} = \ket{b,a}$. This shift is commonly used because it is naturally defined on nonlattice and irregular graphs \cite{Wong13}, and it is necessary for fast search algorithms \cite{AKR2005}.

Coined quantum walks with Grover's diffusion coin and the flip-flop shift have been explored on a variety of graphs, such as the hypercube \cite{Moore2002,Marquezino2008}, 2D grid \cite{Marquezino2010}, and regular graphs \cite{Wong23}. Later, we will give many more examples in the context of searching.

Now we prove that Szegedy's quantum walk is equivalent to this coined quantum walk, with one application of Szegedy's equal to two applications of the coined, \textit{i.e.}, $W = U^2$. Although this was already known in Fact 3.2 of \cite{Magniez2012}, our proof includes the precise relationships between the individual operators composing the walks, and these details are necessary to later prove that Santos's walks are also equivalent to coined quantum walks.

We begin by noting that the Hilbert spaces of the two quantum walks are identical. Recall that Szegedy's quantum walk evolves in $\mathbb{C}^{2|E|}$, and the coined quantum walk evolves in $\mathbb{C}^{\sum_v \text{deg}(v)}$. In any graph, however, the sum of the degrees of the vertices is equal to twice the number of edges, so $\sum_v \text{deg}(v) = 2|E|$. This can be seen in Fig.~\ref{fig:graph_coined}, since each edge supports two directions. For example, the edge connecting vertices $1$ and $2$ supports two basis states, one for a particle at $1$ pointing to $2$ and another at $2$ pointing to $1$. Thus, $\mathbb{C}^{2|E|} = \mathbb{C}^{\sum_v \text{deg}(v)}$.

Now we give a bijection between basis states in Szegedy's quantum walk and the coined quantum walk: Szegedy's walker on the edge connecting vertices $i \in X$ with $j \in Y$ is equivalent to a coined particle at vertex $i$ pointing towards vertex $j$. Both of these states are denoted by the same basis vector $\ket{i,j}$. Thus, the undirected edges in Fig.~\ref{fig:graph_szegedy} and the directed edges in Fig.~\ref{fig:graph_coined} depict the same quantum state.

This bijection allows us to reinterpret Szegedy's quantum walk $W = R_2 R_1$ from its original description on the edges of the bipartite double cover (\textit{c.f.}, Fig.~\ref{fig:graph_szegedy}) to the coined description on the directed graph (\textit{c.f.}, Fig.~\ref{fig:graph_coined}). Let us begin with $R_1$. Recall that in the bipartite double cover, $R_1$ goes through each vertex in $X$ and inverts its edges about their average at the vertex. Then in the language of the coined quantum walk, $R_1$ goes through each vertex in the directed graph, inverting its \emph{outgoing} amplitudes about their average at the vertex. These outgoing amplitudes are the coin states at each vertex, so this is precisely the Grover diffusion coin $C$, so 
\[ R_1 = C. \]

Now consider $R_2$. In the bipartite double cover, $R_2$ goes through each vertex in $Y$ and inverts its edges about their average at the vertex. Reinterpreting this as a coined quantum walk, $R_2$ goes through each vertex of the directed graph and inverts its \emph{incoming} amplitudes about their average at the vertex. This is equivalent to the flip-flop shift followed by the Grover coin and another flip-flop shift, \textit{i.e.},
\[ R_2 = SCS. \]
The first flip-flop shift exchanges incoming and outgoing amplitudes. Then the Grover coin inverts outgoing amplitudes about their mean. Finally, the flip-flop shift again swaps incoming and outgoing amplitudes, so the net effect is that incoming amplitudes, not outgoing amplitudes, were inverted about their means.

\begin{table}
	\caption{\label{table:summary} Summary of the equivalences of Szegedy's and coined quantum walks.}
	\begin{center}
	\begin{tabular}{ccc}
		\hline\noalign{\smallskip}
		\textbf{Szegedy's} & \textbf{Coined} & \textbf{Equivalence} \\
		\noalign{\smallskip}\hline\noalign{\smallskip}
		$W = R_2 R_1$ & $U = SC$ & $W = U^2$ \\
		$W' = R_2' R_1'$ & $U_{\rm SKW} = S(C\ {\rm unmarked}, -I\ {\rm marked})$ & $W' = U_{\rm SKW}^2$ \\
		$W_{q1} = R_2 Q_2 R_1 Q_1$ & $SCQ$ & $W_{q1} = (SCQ)^2$ \\
		$W_{q2} = R_2 R_1 Q_1$ & $U^2Q = SCSCQ$ & $W_{q2} = U^2Q$ \\
		\noalign{\smallskip}\hline
	\end{tabular}
	\end{center}
\end{table}

Combining these two results $R_1 = C$ and $R_2 = SCS$, we have
\[ W = R_2 R_1 = SCSC = U^2, \]
thus proving that two applications of the coined quantum walk is exactly equivalent to one application of Szegedy's quantum walk. This equivalence summarized in the first body row (not the titular row) of Table~\ref{table:summary}.


\subsection{Search with Absorbing Vertices}

Now we give the seminal algorithm by Shenvi, Kempe, and Whaley (SKW) \cite{SKW2003} for searching using a coined quantum walk. In this scheme, the Grover diffusion coin $C$ is still used for unmarked vertices, but the negative identity operator $-I$ is used as the coin at marked vertices. That is, we repeatedly apply the following operator:
\[ U_{\rm SKW} = S \cdot \left\{ \begin{array}{rl}
	C & {\rm on\ unmarked\ vertices} \\
	-I & {\rm on\ marked\ vertices} \\
\end{array} \right\} . \]
This selective coin operator acts as a query to an oracle, and $S$ is the usual flip-flop shift.

Search using SKW's scheme has been explored on a large number of graphs, including the hypercube \cite{SKW2003}, two- and higher-dimensional grids \cite{AKR2005,AR2008,NR2016a}, Sierpinski gaskets \cite{Lara2013}, complete graphs with and without self-loops \cite{Wong10}, and the simplex of complete graphs with a fully marked clique \cite{Wong11}. Other explorations include the impact of internal-state measurements \cite{Wong18} and stationary states \cite{Wong24}.

Now we prove that Szegedy's quantum walk with absorbing vertices is equivalent to SKW's coined quantum walk search scheme. Although this is encompassed in Fact 3.2 of \cite{Magniez2012}, our simple argument gives the reason why.

Recall that in Szegedy's quantum walk with absorbing marked vertices, $R_1'$ and $R_2'$ perform an ``inversion about the mean'' at edges incident to unmarked vertices. So for unmarked vertices, we still have $R_1' = C$ and $R_2' = SCS$. For marked vertices, however, $R_1'$ and $R_2'$ flip the signs of incident edges, so this is equivalent to multiplying them by $-I$. So at unmarked vertices, we have $R_1' = -I$ and $R_2' = S(-I)S$. Thus,
\begin{gather*}
	R_1' = \left\{ \begin{array}{rl}
		C & {\rm on\ unmarked\ vertices} \\
		-I & {\rm on\ marked\ vertices} \\
	\end{array} \right\}, \\
	R_2' = S \left\{ \begin{array}{rl}
		C & {\rm on\ unmarked\ vertices} \\
		-I & {\rm on\ marked\ vertices} \\
	\end{array} \right\} S.
\end{gather*}
Combining these, we see that one application of Szegedy's quantum walk with absorbing marked vertices $W' = R_2' R_1'$ is equivalent to two applications of the coined quantum walk in SKW's search model, \textit{i.e.},
\[ W' = U_{\rm SKW}^2. \]
This is summarized in the second body row of Table~\ref{table:summary}.

Note this equivalence holds no matter how many marked vertices are present. Fact 3.2 of \cite{Magniez2012}, however, only addresses the case of a single marked vertex $w$, for which $U_\text{SKW}$ can be written as \cite{AKR2005}
\[ U_{SKW} = SC (I - 2 \ketbra{w,\phi_w}{w,\phi_w}). \]
The term in parenthesis is a reflection in the Hilbert space, and this is a requirement for Fact 3.2 of \cite{Magniez2012} to apply. As we will see next, the Grover-type queries used in Santos's algorithms do not have this form, and so our upcoming equivalence results lie beyond the scope of Fact 3.2 of \cite{Magniez2012}.


\subsection{Search with Grover's Oracle}

Thus far, we have proved the equivalence of Szegedy's quantum walk and the coined quantum walk, including their seminal search algorithms. Although the general equivalence was already known from Fact 3.2 of \cite{Magniez2012}, our details included the specific relationships between the operators and hold for multiple marked vertices. Now we use these relationships to obtain the coined quantum walks that are equivalent to Santos's search algorithms $W_{q1}$ and $W_{q2}$, which are based on Szegedy's quantum walk with a Grover-type oracle. 

For the coined quantum walk, the Grover-type oracle $Q$ flips the amplitudes at marked vertices. So if the particle is at vertex $a$ pointing towards vertex $b$, the oracle acts as:
\[ Q \ket{a,b} = \left\{ \begin{array}{rl}
	-\ket{a,b}, & a\ {\rm marked} \\
	\ket{a,b}, & a\ {\rm unmarked} \\
\end{array} \right.. \]
This oracle query is \emph{not} a reflection in the full Hilbert space. In the case of a regular graph, the full Hilbert space is the tensor product of vertex and coin spaces, and $Q$ is only a reflection in the vertex space, not the full space. Thus, Grover-type oracles are generally not within the scope of Fact 3.2 of \cite{Magniez2012}.

Now let us consider Santos's first algorithm, $W_{q1} = R_2 Q_2 R_1 Q_1$. We already have $R_1 = C$ and $R_2 = SCS$, so we need to determine $Q_1$ and $Q_2$. Beginning with $Q_1$, recall that in the bipartite double cover (\textit{c.f.}, Fig.~\ref{fig:graph_szegedy_marked_dropped}) that $Q_1$ flips the signs of the edges incident to marked vertices in $X$. By our bijection, this is equivalent to a coined quantum walk (\textit{c.f.}, Fig.~\ref{fig:graph_coined_marked}) where we flip the \emph{outgoing} edges from marked vertices. This is exactly $Q$, so we have 
\[ Q_1 = Q. \]

Now for $Q_2$, in the bipartite double cover, $Q_2$ flips the signs of the edges incident to marked vertices in $Y$. Then in the coined quantum walk, this is equivalent to flipping the signs of \emph{incoming} edges to marked vertices. To implement this, we can apply the flip-flop shift $S$ to swap incoming and outgoing edges, apply the query $Q$ to flip the signs of outgoing edges from marked vertices, and apply the flip-flop shift $S$ again so that the net effect is that the incoming edges to marked vertices had their signs flipped. That is,
\[ Q_2 = SQS. \]

Utilizing these relations, Santos's first search algorithm is $W_{q1} = SCSSQSCQ$. Since $S^2 = I$, this simplifies to
\[ W_{q1} = SCQSCQ = (SCQ)^2. \]
Note that $SCQ$ is a coined quantum walk that takes one walk-step $U = SC$ per oracle query $Q$, and Santos's first algorithm is equivalent to two iterations of it. This is summarized in the third body row of Table~\ref{table:summary}.

This coined quantum walk $SCQ$ is another well-established scheme for searching using a coined quantum walk. It first appeared in \cite{AKR2005}, where this search scheme on the complete graph with a self-loop at each vertex is exactly equivalent to Grover's algorithm (apart from a factor of $2$). Other investigations of this quantum walk search operator $SCQ$ include the complete graph with an arbitrary number of self-loops per vertex \cite{Wong10}, search with potential barriers \cite{Wong13}, improving the success probability by measuring the coin state \cite{Wong18}, and the effect of stationary states \cite{Wong24}. All of these results directly map to Santos's walk $W_{q1}$.

Santos showed that her walk $W_{q1}$ is equivalent to Szegedy's quantum walk with absorbing marked vertices $W'$ when searching strongly regular graphs with a unique marked vertex. As coined quantum walks, this is an equivalence between $SCQ$ and SKW's search algorithm with a selective coin $U_{\rm SKW}$, and it is well understood: If the neighbors of each marked vertex evolve identically due to the symmetry of the graph, then the two operators $SCQ$ and $U_{\rm SKW}$ are identical. This is true for any distance-transitive graph with a unique marked vertex, which includes the complete graph and Johnson graphs in general \cite{Wong20}, the strongly regular graphs in Santos's analysis \cite{Santos2016,Wong5}, the hypercube \cite{SKW2003}, and arbitrary-dimensional lattices \cite{AKR2005}. They can also be equivalent with multiple marked vertices, such as the hypercube with two marked vertices at opposite ``ends'' of the hypercube \cite{Wong18} or the 2D grid with a marked diagonal \cite{AR2008}. This resolves an open question as to when Szegedy's search algorithm can be interpreted as a Grover-type query algorithm.

Now for Santos's second algorithm $W_{q2} = R_2 R_1 Q_1$. Using our previous relations, we have $W_{q2} = SCSCQ$. Since a step of the coined quantum walk is $U = SC$, this yields
\[ W_{q2} = U^2 Q. \]
So Santos's second algorithm is equivalent to a coined quantum walk that takes two walk-steps $U$ for each oracle query $Q$. This is summarized in the last row of Table~\ref{table:summary}.

Taking multiple walk-steps per oracle query is also an existing concept in coined quantum walks. It was introduced by Wong and Ambainis \cite{Wong11} for searching the ``simplex of complete graphs'' for a fully marked cluster. They showed that using multiple walk-steps per oracle query results in an algorithm that is quadratically faster in the number of queries than one that takes one walk-step per oracle query. Santos's result, then, reveals that taking multiple walk-steps per oracle query also improves search on the complete graph, suggesting it could be a general method for speeding up quantum search algorithms.


\section{Discussion and Conclusion}

Szegedy's quantum walk and coined quantum walks are two popular quantum analogues of classical random walks. Szegedy's quantum walk is given by two reflections, while the coined quantum walk is given by a coin flip followed by a shift. Although it is known that one step of Szegedy's quantum walk is equivalent to two steps of the coined, we determined the precise relationship between the operators. In particular, we proved that the first reflection in Szegedy's quantum walk is equivalent to the Grover diffusion coin, while the second reflection is equivalent to the flip-flop shift, Grover diffusion coin, and flip-flop shift again.

These identifications allowed us to reinterpret Santos's two Szegedy-based search schemes with Grover-type oracles as coined quantum walks. As a result of our new equivalence results, a vast literature on coined quantum walks also applies to Santos's schemes. This allowed us to characterize when Szegedy's search algorithm can be expressed in terms of a Grover-type oracle. Before, it was only known for strongly regular graphs with a unique marked vertex, but it is now known for a host of other graphs as well, including graphs with multiple marked vertices. We also showed that Santos's second algorithm is equivalent to searching with multiple steps of the quantum walk per oracle query. This supports the hope that other problems can be similarly sped up using this technique.


\begin{acknowledgements}
	Thanks to Peter H\o{}yer for pointing out Fact 3.2 of \cite{Magniez2012}. This work was supported by the U.S.~Department of Defense Vannevar Bush Faculty Fellowship of Scott Aaronson.
\end{acknowledgements}


\bibliographystyle{qinp}
\bibliography{refs}

\end{document}